\pdfoutput=1

\documentclass[11pt,twoside,a4paper,cmspaper,final,collab]{cms-tdr}
\def\svnVersion{476691P}\def\cmsCernNoTag{CERN-EP-2018-036}\def\cmsCernDate{\today}\def\cmsMessage{Published in the European Physical Journal C as \href{http://dx.doi.org/10.1140/epjc/s10052-018-6242-x}{\doi{10.1140/epjc/s10052-018-6242-x.}}}

\begin{document}\cmsNoteHeader{EXO-16-046}

\hyphenation{had-ron-i-za-tion}
\hyphenation{cal-or-i-me-ter}
\hyphenation{de-vices}
\RCS$HeadURL: svn+ssh://svn.cern.ch/reps/tdr2/papers/EXO-16-046/trunk/EXO-16-046.tex $
\RCS$Id: EXO-16-046.tex 474192 2018-09-07 00:22:37Z apana $

\newlength\cmsTabSkip\setlength{\cmsTabSkip}{1ex}

\newlength\cmsFigWidth
\ifthenelse{\boolean{cms@external}}{\setlength\cmsFigWidth{0.4\textwidth}}{\setlength\cmsFigWidth{0.45\textwidth}}
\newlength\FigThreeWidth
\ifthenelse{\boolean{cms@external}}{\setlength\FigThreeWidth{0.49\textwidth}}{\setlength\FigThreeWidth{0.65\textwidth}}
\newlength\FigFourWidth
\ifthenelse{\boolean{cms@external}}{\setlength\FigFourWidth{0.99\columnwidth}}{\setlength\FigFourWidth{0.7\textwidth}}

\providecommand{\CL}{CL\xspace}

\newcommand{\CHI}{\ensuremath{\chi_{\text{dijet}}}\xspace}
\newcommand{\YBOOST}{\ensuremath{y_{\text{boost}}}\xspace}
\newcommand{\intlumi} {{35.9\fbinv}\xspace}
\newcommand{\mjj} {\ensuremath{M_{\mathrm{jj}}}\xspace}
\newcommand{\ptave}{\ensuremath{\langle p_\mathrm{T}  \rangle}\xspace}
\newcommand{\ptavens}{\ensuremath{\langle p_\mathrm{T}  \rangle}}
\newcommand{\QBHMC} {{\textsc{qbh}}\xspace}
\newcommand{\MADDM} {{\textsc{MadDM}}\xspace}
\newcommand{\QCOUP}{\ensuremath{g_{\mathrm{\cPq}}}\xspace}
\newcommand{\DMCOUP}{\ensuremath{g_{\mathrm{DM}}}\xspace}
\newcommand{\MEDIMASS}{\ensuremath{M_{\text{Med}}}\xspace}
\newcommand{\QBHMASS}{\ensuremath{M_{\mathrm{QBH}}}\xspace}
\newcommand{\DMMASS}{\ensuremath{m_{\mathrm{DM}}}\xspace}
\newcommand{\NEXTRA}{\ensuremath{n_{\mathrm{ED}}}\xspace}
\newcommand{\MCUTOFF}{\ensuremath{M_{\mathrm{S}}}\xspace}
\newcommand{\ETALL}{\ensuremath{\eta_{\mathrm{LL}}}\xspace}
\newcommand{\ETARR}{\ensuremath{\eta_{\mathrm{RR}}}\xspace}
\newcommand{\ETARL}{\ensuremath{\eta_{\mathrm{RL}}}\xspace}
\newcommand{\LAMBDALL}{\ensuremath{\Lambda^{\pm}_{\mathrm{LL}}}\xspace}
\newcommand{\LAMBDARR}{\ensuremath{\Lambda^{\pm}_{\mathrm{RR}}}\xspace}
\newcommand{\LAMBDAVV}{\ensuremath{\Lambda^{\pm}_{\mathrm{VV}}}\xspace}
\newcommand{\LAMBDAAA}{\ensuremath{\Lambda^{\pm}_{\mathrm{AA}}}\xspace}
\newcommand{\LAMBDAVA}{\ensuremath{\Lambda^{\pm}_{(\mathrm{V-A})}}\xspace}
\newcommand{\RelWidth}{\ensuremath{\Gamma/M_{\mathrm{Med}}}\xspace}

\newcolumntype{d}[1]{D{.}{.}{#1}}
\newcolumntype{+}{D{+}{\pm}}

\cmsNoteHeader{EXO-16-046}
\title{Search for new physics in dijet angular distributions using proton-proton collisions
at $\sqrt{s}=13\TeV$ and constraints on dark matter and other models}

\titlerunning{Search for new physics with dijet angular distributions at $\sqrt{s}=13\TeV$}

\author{CMS Collaboration}

\date{\today}

\abstract{
A search is presented for physics beyond the standard model, based on measurements
of dijet angular distributions in proton-proton collisions at $\sqrt{s}=13\TeV$.
The data collected with the CMS detector at the LHC correspond to an integrated
luminosity of \intlumi.
The observed distributions, corrected to particle level, are found to be in agreement with
predictions from perturbative quantum chromodynamics that include electroweak corrections.
Constraints are placed on models containing quark contact
interactions, extra spatial dimensions, quantum black holes, or dark matter, using the
detector-level distributions. In a benchmark model
where only left-handed quarks participate, contact interactions are excluded at the 95\% confidence
level up to a scale of 12.8 or 17.5\TeV, for destructive or constructive interference,
respectively. The most stringent lower limits to date are set on the ultraviolet cutoff in the
Arkani--Hamed--Dimopoulos--Dvali model of extra dimensions.  In the Giudice--Rattazzi--Wells
convention, the cutoff scale is excluded up to 10.1\TeV.  The production of quantum black holes is
excluded for masses below 5.9 and 8.2\TeV, depending on the model.  For the first time, lower limits
between 2.0 and 4.6\TeV are set on the mass of a dark matter mediator for (axial-)vector mediators,
for the universal quark coupling $\QCOUP=1.0$.  }

\hypersetup{
pdfauthor={CMS Collaboration},%
pdftitle={Search for new physics with dijet angular distributions in proton-proton collisions at sqrt(s) = 13 TeV and constraints on dark matter and other models},%
pdfsubject={CMS},%
pdfkeywords={CMS, physics, QCD, contact interactions, extra dimensions, black holes, dark matter}}

\maketitle

\section{Introduction}

{\tolerance=1200
Pairs of highly energetic jets (dijets) are produced at high rates in proton-proton
collisions at the CERN LHC through pointlike scattering of quarks and gluons.  Despite its enormous
success, the shortcomings of the standard model (SM) are well known.  Many theories of physics beyond the
standard model (BSM) that alter the interaction of quarks and gluons from that predicted by
perturbative quantum chromodynamics (QCD) give rise to narrow or wide resonances or even to nonresonant
dijet signatures. Examples that have received widespread attention include models with dark matter
(DM)~\cite{Dreiner:2013vla,Abdallah:2015ter,Boveia:2016mrp, Albert:2017onk, Abercrombie:2015wmb}, quark compositeness~\cite{Terazawa:1980,CI-1,CI-2},
extra spatial dimensions~\cite{ADD1,ADD2}, and quantum black
holes~\cite{BH1,Banks:1999gd,Emparan:2000rs,BH2,BH3}.  Resonances with an intrinsic width of the order of the
experimental resolution can be constrained by searches in the dijet invariant mass
spectrum~\cite{Chala:2015ama,Aaboud:2017yvp,Sirunyan:2016iap}.  These searches, however, are not
very sensitive to wide resonances or nonresonant signatures; a more effective strategy to
constrain such signatures is the study of dijet angular distributions~\cite{Arnison:1986np}.
\par}

The angular distribution of dijets relative to the beam direction is sensitive to the dynamics of
the scattering process. Furthermore, since the angular distributions of the dominant underlying QCD
processes of $\cPq\cPg \to \cPq\cPg$, $\cPq\cPq' \to \cPq\cPq'$, $\cPq\cPq \to \cPq\cPq$,
$\cPg\cPg \to \cPg\cPg$, are all similar~\cite{Combridge:1983jn}, the dijet angular distribution is insensitive
to uncertainties in the parton distribution functions (PDFs). The dijet angular distribution is
typically expressed in terms of $\CHI= \exp(\abs{y_1 -y_2})$, where $y_1$ and $y_2$ are the rapidities
of the two jets with the highest transverse momentum \pt (the leading jets).  For collinear massless parton
scattering, \CHI takes the form $\CHI=(1+\abs{\cos\theta^{*}})/(1-\abs{\cos\theta^{*}})$, where $\theta^{*}$
is the polar scattering angle in the parton-parton center-of-mass (CM) frame.  The choice
of \CHI, rather than $\theta^{*}$, to measure the dijet angular distribution is motivated by the
fact that in Rutherford scattering, where only $t$-channel scattering contributes to the partonic
cross section, the \CHI distribution is independent of $\abs{y_1 -y_2}$~\cite{Combridge:1983jn}.
In contrast, BSM processes may have scattering angle distributions that are closer to being isotropic
than those given by QCD processes and can be identified by an excess of events at small values
of \CHI.  Previous measurements of dijet angular distributions at the LHC have been reported by the
ATLAS~\cite{bib_atlas-chi,bib_atlas-dijet2010,bib_atlas-dijet2011,ATLASDijet8TeV,ATLAS:2015nsi,Aaboud:2017yvp} and
CMS~\cite{bib_cms-chi2010,bib_cms-chi2011,dijetangularCMS8TeV,Sirunyan:2017ygf} Collaborations.

In a simplified model of interactions between DM particles and quarks
~\cite{Bai:2010hh,Shoemaker:2011vi,Dreiner:2013vla,Abdallah:2015ter,Boveia:2016mrp, Albert:2017onk},
the spin-1 (vector or axial-vector) DM mediator particle with unknown
mass \MEDIMASS is assumed to decay only to pairs of quarks or pairs of DM particles, with
mass \DMMASS, and with a universal quark coupling \QCOUP and a DM coupling \DMCOUP. In this model,
the relative width of the DM mediator increases monotonically with increasing \QCOUP. In a scenario
where $\QCOUP=0.25$ and in which the relative widths for vector and axial-vector mediators in the dark
matter decay channels are negligible, values of \MEDIMASS below 3.0\TeV were
excluded by narrow dijet resonance searches~\cite{Aaboud:2017yvp,Sirunyan:2016iap}. 
A search for narrow and broad dijet resonances set constraints on mediator widths up to 30\% ($\QCOUP<0.75$) and 
masses up to 4 TeV~\cite{Sirunyan:2018xlo}. Searches for invisible particles produced in association with quarks or
bosons~\cite{Aaboud:2016tnv,Sirunyan:2017hci,Sirunyan:2017jix} have excluded vector and axial-vector
mediators below 1.8 (2.1)\TeV for $\QCOUP=0.25$ ($\QCOUP=1.0$) and
$\DMCOUP=1.0$~\cite{Sirunyan:2017hci}.

A common signature of quark compositeness~\cite{Terazawa:1980,CI-1,CI-2}, at energies well below the
characteristic mass scale $\Lambda$ for new interactions between quark constituents, is the four-fermion
contact interaction (CI).  The most stringent limits on quark CIs come from searches in dijet
angular distributions at large dijet invariant masses (\mjj)~\cite{Aaboud:2017yvp,Sirunyan:2017ygf},
and in inclusive jet \pt distributions~\cite{jetptCMS}.  The publication from the ATLAS
Collaboration~\cite{Aaboud:2017yvp} provides lower limits on the quark CI scales from 13.1 to
29.5\TeV, depending on the details of the model.

The Arkani--Hamed--Dimopoulos--Dvali (ADD) model \cite{ADD1,ADD2} of compactified large extra
dimensions (EDs) provides a possible solution to the hierarchy problem of the standard model.
It predicts signatures of virtual graviton exchange that result in a nonresonant enhancement of
dijet production in proton-proton collisions, whose angular distribution differs from the
predictions of QCD.  Signatures from virtual graviton exchange have previously been sought at the
LHC in various final states, where the most stringent limits arise from the CMS search with dijet
angular distributions~\cite{Sirunyan:2017ygf}, which excludes the ultraviolet cutoff in the ADD
framework up to 7.9--11.2\TeV, depending on the parameterization of the model.

In models with large EDs, the fundamental Planck scale (\Mpl) is assumed to be closer to the electroweak
(EW) scale, thereby allowing black hole production at the
LHC~\cite{BH1,Banks:1999gd,Emparan:2000rs,BH2,BH3}.  Semiclassical black holes, which have mass much
larger than \Mpl, decay into multiple jets through Hawking radiation~\cite{BH4}.
Quantum black holes (QBHs), which are produced with mass close to \Mpl, decay predominantly
into dijets and can be studied using dijet angular distributions~\cite{QBH,QBH1,QBH2}. Recent
searches for QBHs with dijet final states at the LHC reported in
Refs.~\cite{Aaboud:2017yvp,Sirunyan:2017ygf} exclude QBHs with masses below 8.9\TeV.

In this paper, we present a search for new physics, specifically DM mediators, CIs, EDs, and QBHs,
using measurements of dijet angular distributions.
The signature of the signals can be categorized into nonresonant
excesses at high \mjj as predicted by the CI and ADD models and resonances from the decay of QBHs
and DM mediators that could appear across the whole range of the \mjj spectrum.
The searches are performed by comparing detector-level
dijet angular distributions with BSM predictions that have been adjusted to include detector resolution
effects. This eliminates some systematic uncertainties that are introduced when correcting the dijet
angular distributions for detector effects and simplifies the statistical evaluation.
The dijet angular distributions are
also corrected to particle level to facilitate comparisons with other theoretical predictions
and published in HEPData.

\section{The CMS detector}

{\tolerance=1200
The CMS apparatus is based on a superconducting solenoid of 6\unit{m} internal diameter, providing
an axial field of 3.8\unit{T}. Within the solenoid and nearest to the interaction point are the
silicon pixel and strip trackers.  Surrounding the tracker volume are the lead tungstate crystal
electromagnetic calorimeter and the brass and scintillator hadron calorimeter.  The trackers cover a
pseudorapidity region of $\abs{\eta} < 2.5$ while the calorimeters cover $\abs{\eta} < 3.0$.  In
addition, CMS has extensive forward calorimetry, which extends the coverage to $\abs{\eta}<5.0$.
Finally, muons are measured in gas-ionization detectors embedded in the steel flux-return yoke of
the solenoid, with a coverage of $\abs{\eta} < 2.4$.  A two-tiered system, with a level-1 trigger
followed by a high-level trigger (HLT), is used by CMS to record events of
interest~\cite{Khachatryan:2016bia} for the offline analysis.  A more detailed description of the
CMS detector, together with a definition of the coordinate system used and the relevant kinematic
variables, can be found in Ref.~\cite{bib_CMS}.
\par}

\section{Event selection and data unfolding}

Events are reconstructed using a particle-flow algorithm \cite{Sirunyan:2017ulk} to identify and
reconstruct individual particles from each collision by combining information from all CMS
subdetectors.  Identified particles include charged and neutral hadrons, electrons, muons, and
photons.  The particles are clustered into jets using the anti-\kt
algorithm~\cite{antikt:2008,Cacciari:2011ma} with a distance parameter of 0.4.  In order to mitigate
the effect of additional proton-proton interactions within the same or nearby bunch crossings
(pileup) on the jet momentum measurement, the charged hadron subtraction technique~\cite{Sirunyan:2017ulk} is
used.  Spurious jets from noise or non-collision backgrounds are rejected by applying jet
identification criteria~\cite{CMS-PAS-JME-16-003}.  The jet energies are corrected for nonlinear and
nonuniform response of the calorimeters through corrections obtained from data and Monte Carlo (MC)
simulations~\cite{Khachatryan:2016kdb}.  To compare data with theoretical predictions, the same jet
clustering algorithm is applied to the generated stable particles (lifetime $c\tau > 1\cm$) from MC
simulations with leading order (LO) {\PYTHIA 8.212}~\cite{pythia6,pythia8:2008} predictions, and to
the outgoing partons from next-to-leading (NLO) predictions.

The events used in this analysis are selected with triggers based upon either jet \pt or \HT, as measured by the HLT,
where \HT is the scalar sum of the \pt values of all the jets with $\abs{\eta} < 3.0$ and \pt
greater than 30\GeV.  The HLT selection requires having a jet with $\pt>450\GeV$ or an \HT value of
at least 900\GeV.  The data sample was collected with the CMS detector in 2016 and corresponds to an
integrated luminosity of \intlumi~\cite{CMS-PAS-LUM-17-001}.

In the subsequent offline analysis, events with a reconstructed primary vertex that
lies within $\pm$24\unit{cm} of the detector center along the beam line, and within 2\unit{cm} of
the detector center in the plane transverse to the beam, are selected.  The primary vertex is
defined as the reconstructed vertex with the highest sum of the squares of all associated
physics objects \pt.  The physics objects are the jets returned by the application of the anti-\kt algorithm
to all tracks associated with the vertex, plus the corresponding associated missing transverse
momentum, taken as the negative vector sum of the \pt of those jets.

The two leading jets are used to measure the dijet angular distributions in seven regions of the dijet
invariant mass \mjj.  The \mjj regions, in units of \TeV, are chosen to be 2.4--3.0, 3.0--3.6,
3.6--4.2, 4.2--4.8, 4.8--5.4, 5.4--6.0, and $>$6.0.  The highest \mjj range was chosen to maximize
the expected sensitivity to the BSM signals considered.  The phase space for this analysis is restricted
by the requirements $\CHI<16$ and $\abs{\YBOOST}<1.11$, where $\YBOOST = (y_1 + y_2)/2$.  This selection
and the \mjj range definition restrict the absolute rapidities $\abs{y_1}$ and $\abs{y_2}$ of the
two highest \pt jets to be less than 2.5 and their \pt to be larger than 200\GeV.  The trigger
efficiency for events that satisfy the subsequent selection criteria exceeds 99\% in all the \mjj
ranges for the analysis.  The observed numbers of events in the analysis
phase space for each of the mass ranges are 353025, 71832, 16712, 4287, 1153, 330, and 95.  The
highest value of \mjj observed among these events is 8.2\TeV.

{\tolerance=2400
In this paper, we present dijet angular distributions normalized to unity in each
\mjj range, denoted $(1/\sigma_\text{dijet})(\rd\sigma_\text{dijet}/\rd\chi_\text{dijet})$,
where $\sigma_\text{dijet}$ is the cross section in the analysis phase space.
\par}

Fluctuations in jet response from the resolution in jet \pt of the detector can cause lower energy
jets to be misidentified as leading jets and also result in bin-to-bin event migrations in both \CHI
and dijet mass.  The corrections for these effects are obtained from a two-dimensional response
matrix that maps the generator-level \mjj and \CHI distributions onto the measured values. This
matrix is obtained using particle-level jets from the \PYTHIA MC event generator that are smeared
in \pt using a double-sided Crystal Ball parameterization~\cite{crystalball} of the response. This
parameterization takes into account the full jet energy resolution, including non-Gaussian tails,
and is derived from the full detector simulation. The width of the Gaussian core in the
parameterization is adjusted to account for the difference in resolution observed between data and
simulation~\cite{Khachatryan:2016kdb}. The reason for deriving the response matrix from smeared
generator-level MC rather than from full detector simulation is that significantly smaller
statistical uncertainties can be achieved using the faster code.  The measured distributions are
unfolded to particle level by inverting the response matrix without regularization, using
the \textsc{RooUnfold} package~\cite{roounfold}.  The unfolding changes the shape of the \CHI
distributions by $<$1\% and $<$8\% across \CHI in the lowest and highest \mjj ranges, respectively.
The fractions of event migrations between mass bins are 15--20\% in the lowest \mjj range and
25--40\% in the highest \mjj range, depending on \CHI values.
The unfolding procedure was tested by splitting the simulation data into independent training and
testing samples. The training sample was used to derive a response matrix and the
smeared \CHI distributions from the test sample were unfolded using this response matrix.
No significant difference was observed between the generated and unfolded \CHI distributions
in the test sample. The effects of migrations between \CHI bins are negligible.  The unfolding
procedure is based on matrix inversion, while the procedure used in previous
publications of dijet angular distributions~\cite{dijetangularCMS8TeV,Sirunyan:2017ygf} was based on
the D'Agostini iterative method~\cite{D'Agostini1995487}.  We have compared these two methods by
deriving limits from unfolded data, and the limits vary by less than 5\%.

\section{Theoretical predictions}

We compare the unfolded normalized dijet angular distributions with the predictions of perturbative
QCD at NLO, available in \textsc{nlojet++}~4.1.3~\cite{nlojet} in the \textsc{fastnlo}~2.1
framework~\cite{fastnlo}. EW corrections for dijet production~\cite{Dittmaier:2012kx} change the
predicted normalized distributions by up to 1\% (5\%) for the lowest \CHI bins in small (large)
values of \mjj.  The factorization ($\mu_{\mathrm{f}}$) and renormalization ($\mu_{\mathrm{r}}$)
scales are set to the average \PT of the two jets, $\ptave=(\PT{}_{1}+\PT{}_{2})/2$, and
the PDFs are taken from the CT14 set~\cite{Dulat:2015mca}. The use of a more flexible statistical
combination of multiple PDF sets as in PDF4LHC15\_100~\cite{Butterworth:2015oua, Dulat:2015mca,
Harland-Lang:2014zoa, Ball:2014uwa, Gao:2013bia, Carrazza:2015aoa} exhibited small differences as
compared to the CT14 PDF set.  We evaluated the impact of nonperturbative effects from hadronization
and multiple parton interactions on the QCD predictions using \PYTHIA{} with the CUETP8M1
tune~\cite{Skands:2014pea} and {\HERWIGpp} 2.7.1~\cite{herwig:2008} with tune
EE5C~\cite{Seymour:2013qka}. The effects are found to be less than 1\% and negligible for both MC
generators.

The production and decay of the DM mediators in the simplified DM model are generated at LO
using \MADDM version 2.0.6~\cite{Backovic:2013dpa,Backovic:2015cra} at fixed \DMCOUP
and \DMMASS values, where $\DMCOUP = 1.0$ and $\DMMASS = 1\GeV$.  For these values of \DMCOUP
and \DMMASS, the differences between vector and axial-vector mediators in the cross sections and in
the acceptances are negligible in the analysis phase space.

BSM physics signatures from CIs with flavor-diagonal color-singlet couplings among quarks are described
by the effective Lagrangian~\cite{CI-1,CI-2}:
\ifthenelse{\boolean{cms@external}}{
\begin{equation*}
\begin{split}
  \mathcal{L}_{\cPq\cPq}=\frac{2\pi}{\Lambda^2} [
    & \ETALL({\PAQq}_{\mathrm{L}}\gamma^{\mu}\cPq_{\mathrm{L}})(\PAQq_{\mathrm{L}}\gamma_{\mu}\cPq_{\mathrm{L}}) + \\
    & \ETARR(\PAQq_{\mathrm{R}}\gamma^{\mu}\cPq_{\mathrm{R}})(\PAQq_{\mathrm{R}}\gamma_{\mu}\cPq_{\mathrm{R}}) + \\
    & 2\ETARL(\PAQq_{\mathrm{R}}\gamma^{\mu}\cPq_{\mathrm{R}})(\PAQq_{\mathrm{L}}\gamma_{\mu}\cPq_{\mathrm{L}})],
\end{split}
\end{equation*}
}{
\begin{equation*}
  \mathcal{L}_{\cPq\cPq}=\frac{2\pi}{\Lambda^2} \left[
    \ETALL({\PAQq}_{\mathrm{L}}\gamma^{\mu}\cPq_{\mathrm{L}})(\PAQq_{\mathrm{L}}\gamma_{\mu}\cPq_{\mathrm{L}})
    +\ETARR(\PAQq_{\mathrm{R}}\gamma^{\mu}\cPq_{\mathrm{R}})(\PAQq_{\mathrm{R}}\gamma_{\mu}\cPq_{\mathrm{R}})
    +2\ETARL(\PAQq_{\mathrm{R}}\gamma^{\mu}\cPq_{\mathrm{R}})(\PAQq_{\mathrm{L}}\gamma_{\mu}\cPq_{\mathrm{L}})
  \right],
\end{equation*}
}
where the subscripts $\mathrm{L}$ and $\mathrm{R}$ refer to the left and right chiral projections of
the quark fields, respectively, and \ETALL, \ETARR, and \ETARL are taken to be 0, ${+}1$, or ${-}1$
for the different combinations that correspond to different CI models. The following CI
possibilities with color-singlet couplings among quarks are investigated:

\begin{table}[htb!!!]
\centering
    \begin{tabular}{cc}
      Model & $\left(\ETALL,\,\ETARR,\,\ETARL\right)$\\\hline
      \LAMBDALL    & $(\pm 1,\hphantom{\pm}0,\hphantom{\pm}0)\,$\\
      \LAMBDARR    & $(\hphantom{\pm}0,\pm 1,\hphantom{\pm}0)\,$\\
      \LAMBDAVV    & $(\pm 1,\pm 1,\pm 1)\,$\\
      \LAMBDAAA    & $(\pm 1,\pm 1,\mp 1)\,$\\
      \LAMBDAVA & $(\hphantom{\pm}0,\hphantom{\pm}0,\pm 1)\,$\\
    \end{tabular}
\end{table}

The models with positive (negative) \ETALL or \ETARR lead to destructive (constructive) interference
with the QCD terms, and consequently a lower (higher) cross section, respectively.  In all CI models
discussed in this paper, NLO QCD corrections are employed to calculate the cross sections.  In
proton-proton collisions, the \LAMBDALL and
\LAMBDARR models result in identical lowest order cross sections and NLO corrections,
and consequently lead to the same sensitivity.  For \LAMBDAVV and \LAMBDAAA, as well as for
\LAMBDAVA, the CI predictions are also
identical at lowest order, but exhibit different NLO corrections and yield different sensitivities.
The \textsc{cijet}~1.0 program~\cite{contactnlo} is used to calculate the CI terms, as well as the
interference between the CI and QCD terms at NLO in QCD.

For the ADD model, two parameterizations for virtual graviton exchange are considered:
Giudice--Rattazzi--Wells (GRW)~\cite{GRW} and Han--Lykken--Zhang (HLZ)~\cite{HLZ}.  In the GRW
convention, the sum over the Kaluza--Klein graviton excitations in the effective field theory is
regulated by a single cutoff parameter $\Lambda_{\mathrm{T}}$.  In the HLZ convention, the effective
theory is described in terms of two parameters, the cutoff scale \MCUTOFF and the number of extra
spatial dimensions \NEXTRA.  The parameters \MCUTOFF and \NEXTRA are directly related to
$\Lambda_{\mathrm{T}}$~\cite{landsberg}.  We consider models with 2--6 EDs.  The case of $\NEXTRA=1$
is not considered since it would require an ED of the size of the radius of the solar system; the gravitational
potential at such distances would be noticeably modified, and this case is therefore excluded by
observation. The case of $\NEXTRA=2$ is special in the sense that the relation between \MCUTOFF and
$\Lambda_{\mathrm{T}}$ also depends on the parton-parton CM energy $\sqrt{s}$.  The ADD predictions
are calculated using \PYTHIA.

Quantum black hole production is studied within the framework of the ADD model, with $\NEXTRA=6$
(ADD6), and the Randall--Sundrum model (RS1)~\cite{RS1,RS2} with a single, warped extra dimension
($\NEXTRA=1$).  In these models, the QBH production cross section depends on the mass of the QBH,
\Mpl, and the number of spatial dimensions.  Since QBHs are produced with a mass
threshold close to \Mpl, we set the minimum QBH mass \QBHMASS equal to \Mpl for
simplicity.  The \QBHMC 3.0 generator~\cite{QBHGenRef} is used for the predictions.

{\tolerance=2400
To take into account the NLO QCD and EW corrections to SM dijet production when probing the ADD,
QBH, and DM models, the cross section difference $\sigma^{\mathrm{QCD}}_{\text{NLO+EW corr}}
- \sigma^{\mathrm{QCD}}_{\text{LO}}$ is evaluated for each \mjj and \CHI bin and added to the SM+BSM
predictions.  This procedure provides an SM+BSM prediction where the QCD terms are corrected to NLO
with EW corrections while the BSM terms are calculated at LO.  While the ADD BSM prediction from
\PYTHIA  includes the interference terms of graviton exchange with QCD (obtained by subtracting
the predictions $\sigma^{\text{ADD+QCD}}_{\text{LO}} - \sigma^{\text{QCD}}_{\text{LO}}$), the QBH
and DM BSM predictions do not include such interference terms.
\par}

{\tolerance=1200
Exclusion limits on the BSM models studied in this paper are set based on the
comparison of data that have not been corrected for resolution effects with both SM+BSM and SM
predictions that have been folded to detector level. The comparison at detector level is done to
eliminate some systematic uncertainties that are introduced during the unfolding procedure and
simplifies the statistical evaluation. This procedure uses the same two-dimensional response matrix
whose inverse is used for unfolding the data.
It has been verified that the \CHI distributions for SM+BSM predictions
folded with the response matrix derived from SM QCD multijet predictions smeared with the
double-sided Crystal Ball parameterization of the jet \pt resolution agree with SM+BSM predictions
smeared with this same parameterization. The folding procedure is equivalent to running the full
detector simulation on the particle-level predictions, with the residual differences accounted for
in the systematic uncertainties.
\par}

\section{Systematic uncertainties}

The normalized \CHI distributions are relatively insensitive to many potential systematic effects.  To present
the uncertainties for the normalized shapes, the quoted values are reported for
the lowest \CHI bins, where the uncertainties and potential contributions from BSM processes are
typically the largest.
The main experimental uncertainty is from the jet energy scale (JES) and the
main theoretical uncertainty is from the choices of $\mu_{\mathrm{r}}$ and $\mu_{\mathrm{f}}$
scales.

\subsection{Experimental uncertainties}

The overall JES uncertainty is less than 1\%, and the variation of the JES as a function of
pseudorapidity is less than 1\% per unit $\eta$~\cite{Khachatryan:2016kdb,jets2015} in the phase
space of the analysis.  The JES uncertainties related to each step in the derivation of the \pt
and $\eta$ dependent JES corrections are taken into account independently. In this way, the
correlations of the JES uncertainty sources among the \mjj ranges and \CHI bins are included.
For the purpose of display in figures and tables, the total JES uncertainty is obtained from the
quadratic sum of these uncertainty sources and is found to be 3.6\% in the lowest \mjj range and
9.2\% in the highest \mjj range.

The uncertainty from the jet \pt resolution is evaluated by changing the width of the Gaussian core
of the Crystal Ball parameterization of the response by up to
$\pm$5\%~\cite{Khachatryan:2016kdb,jets2015}, depending upon the jet $\eta$, and comparing the
resultant distributions before and after these changes. This uncertainty is found to be less than
1\% for all \mjj.  The uncertainty from the modeling of the tails of the jet \pt resolution~\cite{bib_jecjinst} is evaluated using a
Gaussian function to parameterize the response, and we assign an uncertainty equal to half of the
difference between the distributions determined from this Gaussian ansatz and the nominal
correction. The size of this uncertainty is less than 1.5\% for all \mjj.

Another source of uncertainty arises from the use of a parametric model to simulate the jet \PT
resolution of the detector.  This uncertainty is estimated by comparing the smeared
\CHI distributions to the ones from a detailed
simulation of the CMS detector using \GEANTfour~\cite{Agostinelli:2002hh}, and is found to be 0.5\%
and 1\% in the lowest and highest \mjj ranges, respectively.

{\tolerance=1200
In the unfolding procedure, there is an additional systematic uncertainty introduced due to
potential mismodeling of the dijet kinematic distributions in \PYTHIA. This uncertainty is evaluated
using \MGvATNLO 2.2.2~\cite{Alwall:2014hca} predictions, as the kinematic
distributions from \MGvATNLO and \PYTHIA  are found to bracket the data.  The
inverted response matrix from \PYTHIA  is applied to the smeared \CHI distributions
from \MGvATNLO and the results are compared to the corresponding generated \CHI
distributions.  The differences are observed to be less than 1.5\% for all \mjj.
\par}

The effect from pileup is studied by comparing the \CHI distributions with various numbers of pileup
interactions in simulated events.  The numbers are varied according to the uncertainty of the total
inelastic cross section of $\Pp\Pp$ collisions~\cite{Aaboud:2016mmw}.  The effect on the \CHI
distributions is observed to be negligible.

\subsection{Theoretical uncertainties}

The uncertainties due to the choices of $\mu_{\mathrm{f}}$ and $\mu_{\mathrm{r}}$ scales in the NLO
QCD predictions are evaluated by following the proposal in Refs.~\cite{Cacciari:2003fi,Banfi:2010xy}
and changing the default choice of scales in the following 6 combinations:
$(\mu_{\mathrm{f}}/\ptave$, $\mu_{\mathrm{r}}/\ptave)$ = $(1/2, 1/2)$, $(1/2, 1)$, $(1,1/2)$, $(2,
2)$, $(2, 1)$, and $(1, 2)$.  These changes modify the predictions of the normalized \CHI
distributions by up to 8.5\% and up to 19\%, at small and large values of \mjj, respectively.  The
uncertainty in the NLO QCD predictions due to the choice of PDFs is determined from the 28
eigenvectors of CT14 using the procedure described in Ref.~\cite{cteq}, and is found to be less than
0.2\% at low \mjj and less than 0.6\% at high \mjj. The uncertainty in the strong coupling constant
has a negligible impact on the normalized \CHI distribution.

Scale and PDF uncertainties in the CI predictions are obtained using the same procedure as in the
QCD predictions.  In the ADD and QBH models, the scale and PDF uncertainties have a negligible
impact on the limits as the signals only appear in the highest mass bins, where the statistical
uncertainties dominate. The effect on the acceptance for the DM models due to the PDF uncertainty is
evaluated using the 100 replica NNPDF3.0 PDF set~\cite{Ball:2014uwa} and found to be non-negligible
in the \mjj ranges with $\mjj>\MEDIMASS$ for DM mediators that have large mass and coupling.  For
example, for an axial-vector mediator with $\MEDIMASS = 6\TeV$ and $\QCOUP = 1.0$, which corresponds
to a resonance with relative width of 50\%, the uncertainty is 14\% in the $\mjj>\!6.0\TeV$ bin.

Although the uncertainties are treated separately in the statistical analysis of the data, for
display purposes in tables and figures we calculate the total experimental and theoretical uncertainty as
the quadratic sum of the contributions due to the JES, the jet \PT resolution,
the modeling of both the detector response and the dijet kinematics,
and the contributions from $\mu_{\mathrm{f}}$, $\mu_{\mathrm{r}}$, and the PDFs.  A
summary of the leading experimental systematic uncertainties is provided in Table~\ref{tab:sys}.
The theoretical uncertainties quoted in the table apply to the QCD prediction.
As shown in Table~\ref{tab:sys}, systematic uncertainties dominate the total
uncertainty in low \mjj regions, while the statistical uncertainty dominates in high \mjj regions.

\begin{table*}[htb!]
\centering
\topcaption{\label{tab:sys}
Summary of the leading experimental and theoretical uncertainties in the normalized \CHI distributions,
in percent.  While the statistical analysis represents each uncertainty through a change in the \CHI
distribution correlated among all \CHI bins, this table summarizes each uncertainty by a
representative value to show their relative contributions.  For the lowest and highest
dijet mass bins, the relative shift is given for the lowest \CHI bin.  In the highest dijet mass
bin, the dominant experimental contribution corresponds to the statistical uncertainty, while the
dominant theoretical contribution is from the uncertainty in scales.}
\begin{tabular}{lcc}
Source of uncertainty & $2.4<\mjj<3.0\TeV$ & $\mjj>6.0\TeV$ \\[2pt]
\hline\\[-2ex]
Statistical & 0.7 & 27 \\
JES & 3.6 & 9.2 \\
Jet \PT resolution (core) & 1.0 & 1.0 \\
Jet \PT resolution (tails) & 1.0 & 1.5 \\
Detector response model & 0.5 & 1.0 \\
Unfolding, model dependence & 0.2 & 1.5 \\[\cmsTabSkip]
Total experimental & 4.1 & 29 \\[\cmsTabSkip]
QCD NLO scale (6 changes in $\mu_{\mathrm{r}}$ and $\mu_{\mathrm{f}}$) & $_{-3.0}^{+8.5}$ & $_{-5.8}^{+19}$ \\[1pt]
PDF (CT14 eigenvectors) & 0.2 & 0.6 \\[\cmsTabSkip]
Total theoretical & 8.5 & 19 \\
\end{tabular}
\end{table*}

\section{Results}

\begin{figure*}[htbp]
\centering
\includegraphics[width=\cmsFigWidth]{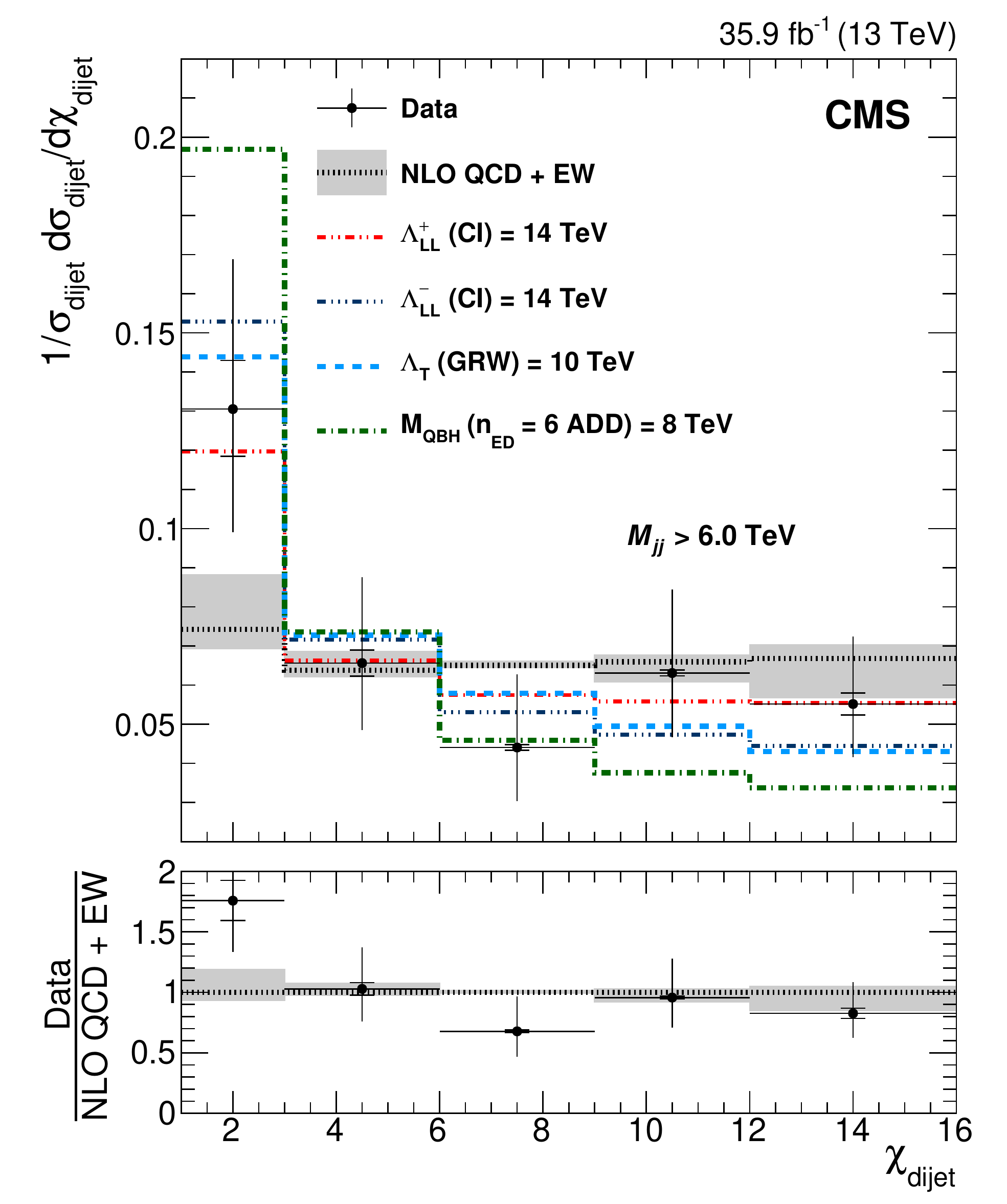}
\includegraphics[width=\cmsFigWidth]{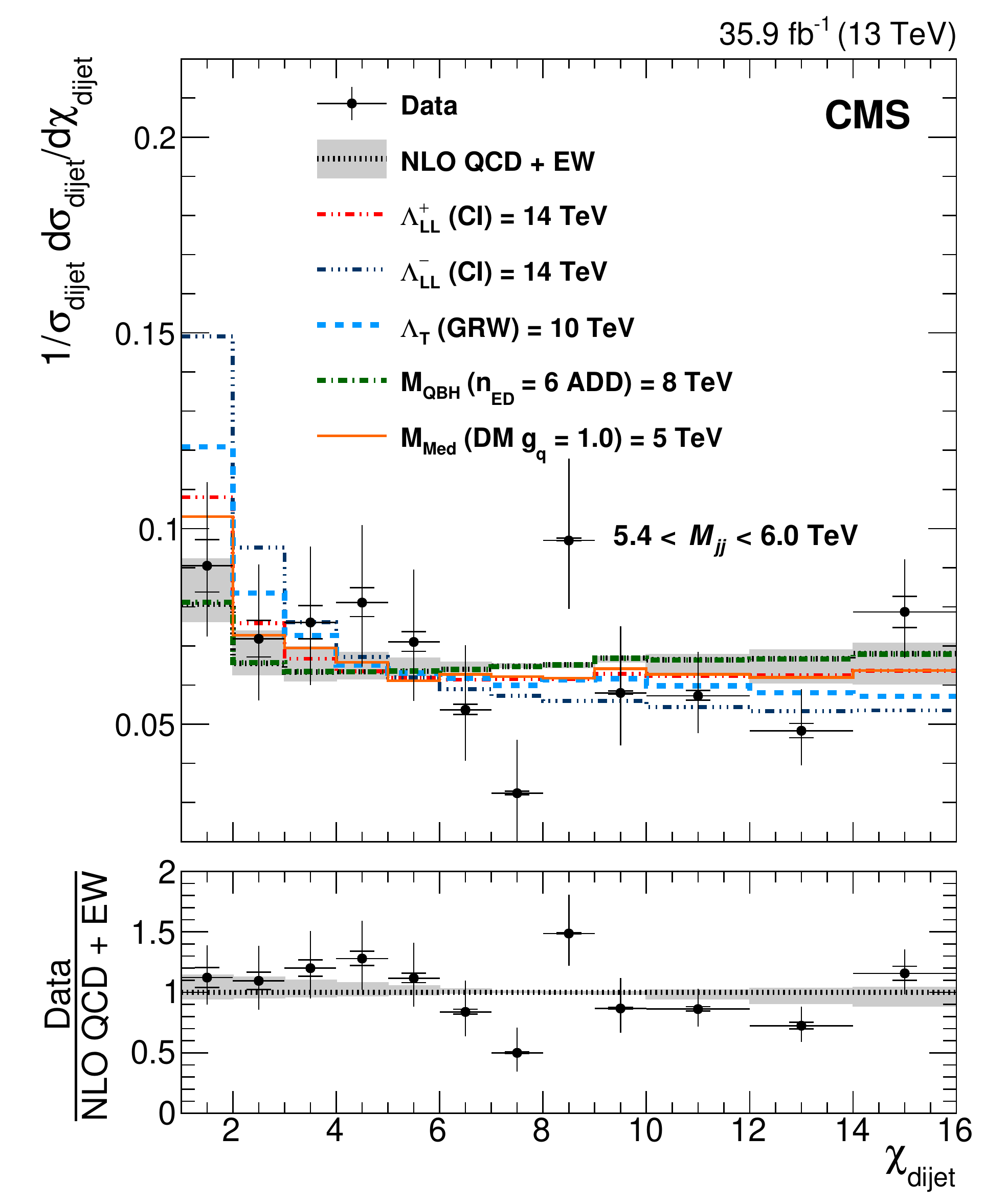}
\includegraphics[width=\cmsFigWidth]{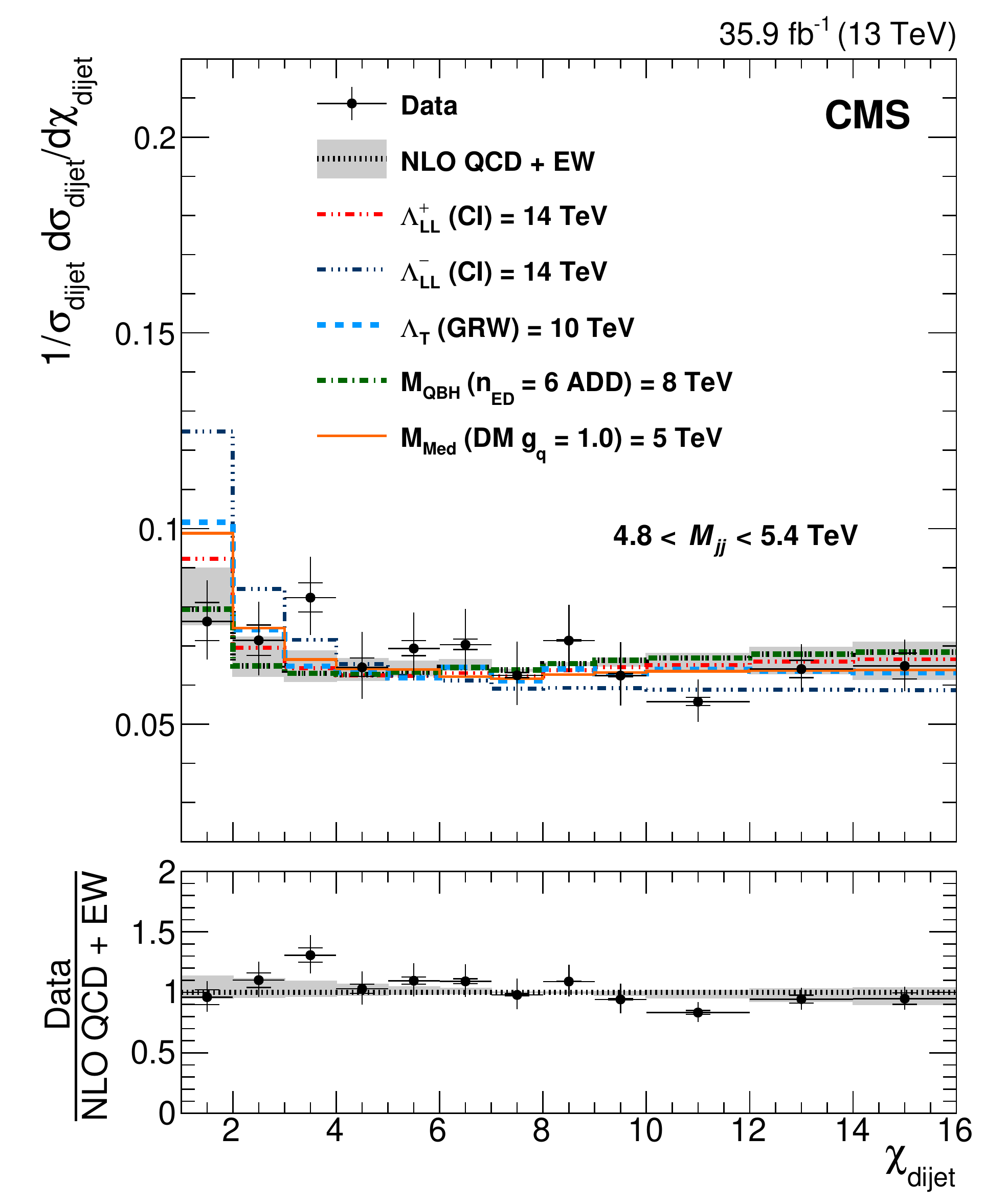}
\caption{Normalized \CHI distributions in the three highest mass bins.
 Unfolded data are compared to NLO predictions (black dotted line).
 The error bars represent statistical and experimental systematic uncertainties combined in quadrature.
 The ticks on the error bars correspond to the experimental systematic uncertainties only.
 Theoretical uncertainties are indicated as a gray band. Also shown are the predictions for
 various CI, ADD, QBH, and DM scenarios.
 The lower panels show the ratio of the unfolded data distributions and NLO predictions.
 }
\label{fig:data_results2}
\end{figure*}

\begin{figure*}[htbp]
\centering
\includegraphics[width=\cmsFigWidth]{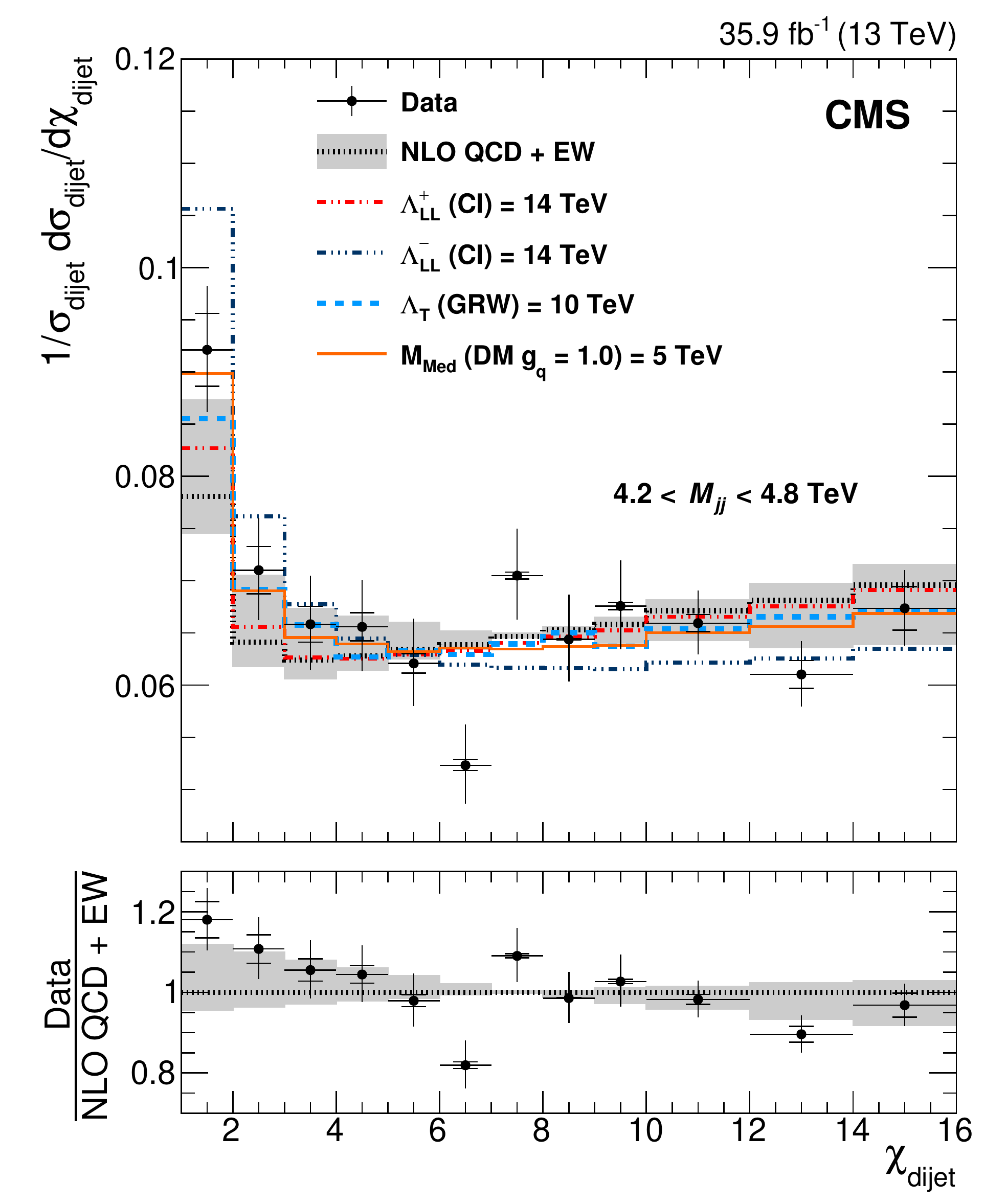}
\includegraphics[width=\cmsFigWidth]{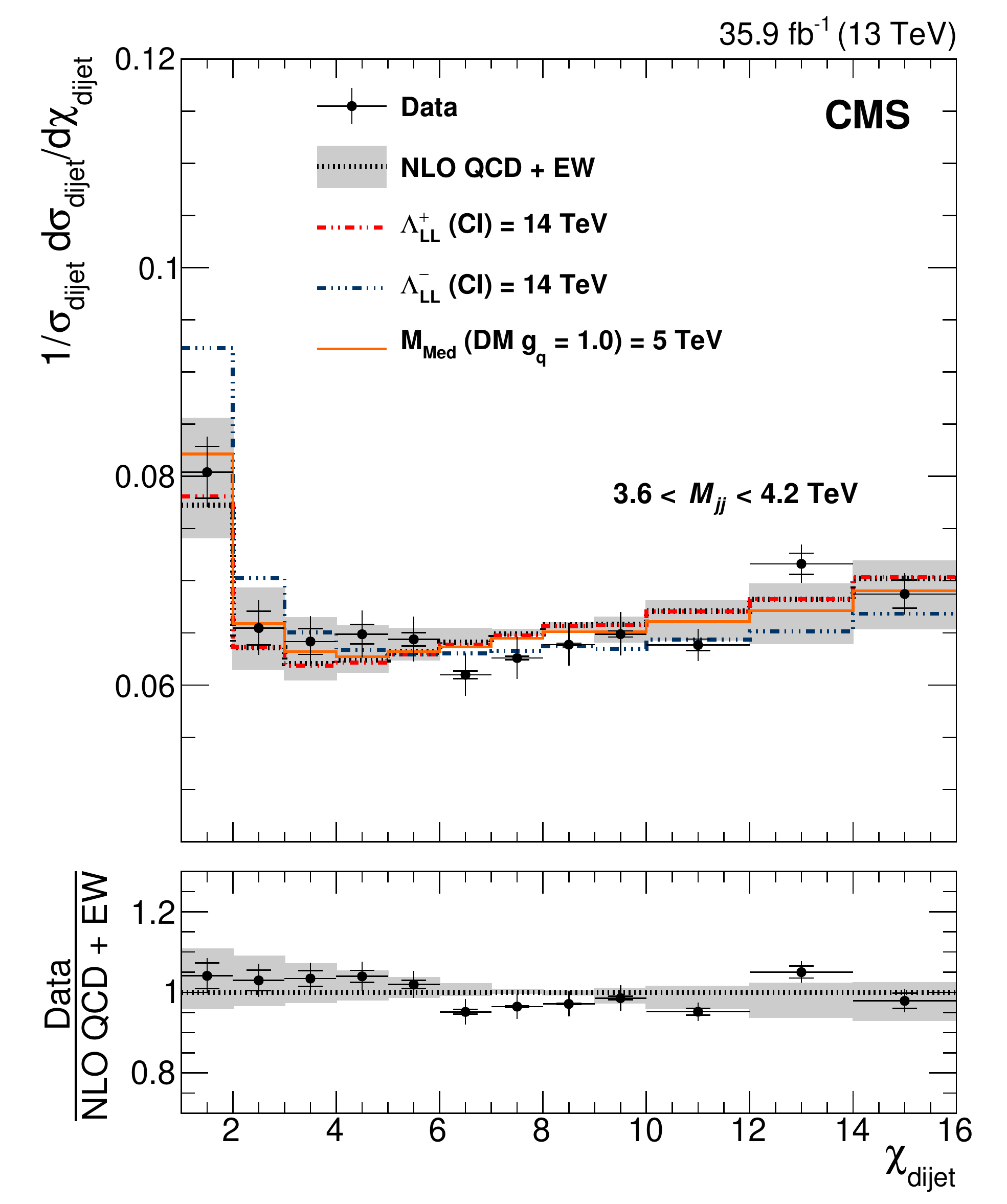}
\includegraphics[width=\cmsFigWidth]{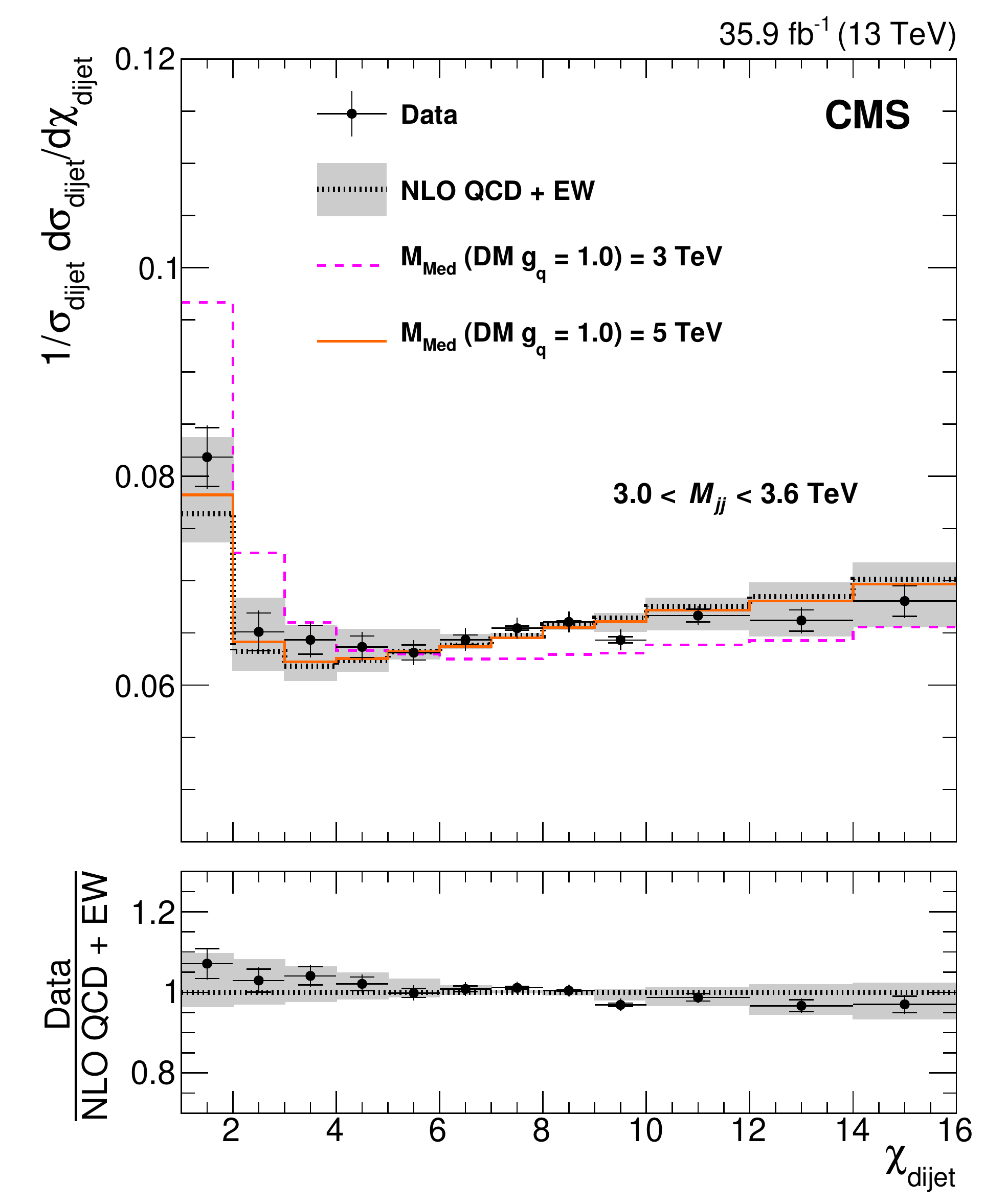}
\includegraphics[width=\cmsFigWidth]{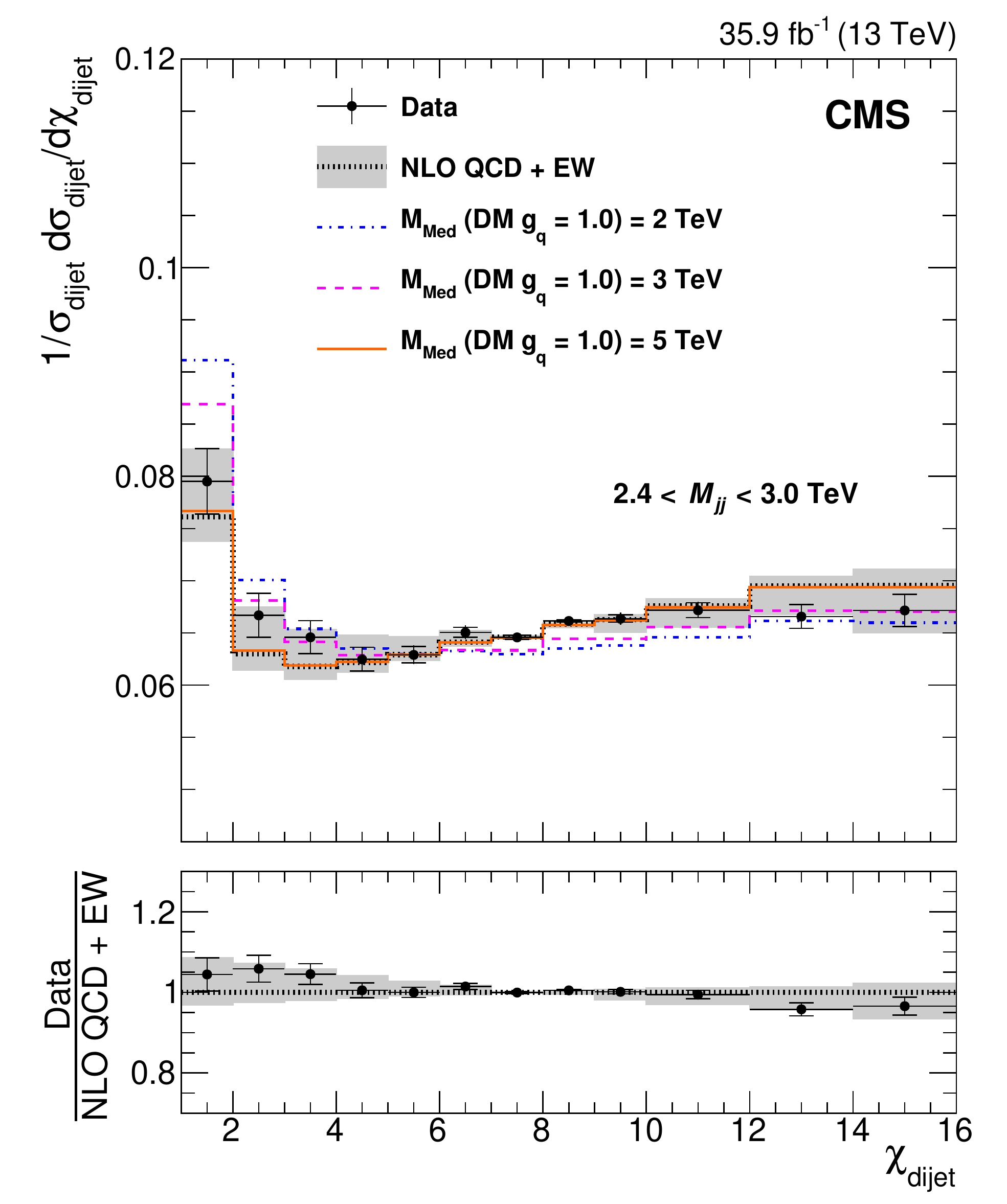}
\caption{Normalized \CHI distributions in the four lower mass bins.
 Unfolded data are compared to NLO predictions (black dotted line).
 The error bars represent statistical and experimental systematic uncertainties combined in quadrature.
 The ticks on the error bars correspond to the experimental systematic uncertainties only.
 Theoretical uncertainties are indicated as a gray band. Also shown are the predictions for
 various CI, ADD, and DM scenarios.
 The lower panels show the ratio of the unfolded data distributions and NLO predictions.
 }
\label{fig:data_results3}
\end{figure*}

{\tolerance=1200
In Figs.~\ref{fig:data_results2}~and \ref{fig:data_results3}
the measured normalized \CHI distributions for all mass bins unfolded to particle level are compared
to NLO predictions with EW corrections. No significant deviation from the SM prediction is observed.  The
distributions are also compared to predictions for QCD+CI with CI scales equal to 14\TeV, QCD+ADD
with $\Lambda_{\mathrm{T}}\ (\mathrm{GRW})\ =10\TeV$, QCD+QBH with $\QBHMASS~(\mathrm{ADD6}) = 8\TeV$, and
QCD+DM with $M_{\mathrm{Med}}=2$, 3 and 5\TeV and $\QCOUP=1.0$.  The signal distributions are only
shown for the \mjj ranges that contribute to the sensitivity for the BSM searches.
\par}

{\tolerance=1200 The asymptotic approximation~\cite{AsymptCLs} of the $\mathrm{\CL_s}$
criterion~\cite{junk:1999,read:2002} is used to set exclusion limits on the parameters for the BSM
models~\cite{CMS-NOTE-2011-005}. The limits obtained using this approximation were tested against the
$\mathrm{\CL_s}$ limits obtained using ensembles of pseudo experiments for several of the models
examined, and the differences were found to be negligible. The likelihoods $L_\mathrm{QCD}$ and
$L_\mathrm{QCD+BSM}$ are defined for the respective QCD-only and QCD+BSM hypotheses as a product of
Poisson likelihood functions for each bin in \CHI.  The predictions for each \mjj range are
normalized to the number of observed events in that range.  Systematic uncertainties are treated as
nuisance parameters in the likelihood model. Following Ref.~\cite{Aaboud:2017yvp}, the nuisance
parameters are profiled with respect to the QCD-only and QCD+BSM models by maximizing the
corresponding likelihood functions.  The $p$-values for the two hypotheses,
$P_\mathrm{QCD+BSM}(q \geq q_\text{obs})$ and $P_\mathrm{QCD}(q \leq q_\text{obs})$, are evaluated
for the profile log-likelihood ratio $q = -2 \ln ({L_\text{QCD+BSM}}/{L_\mathrm{QCD}})$.  Limits on
the QCD+BSM models are set based on the quantity $\mathrm{\CL}_\mathrm{s} = P_\mathrm{QCD+BSM}(q\geq
q_\text{obs}) / (1-P_\mathrm{QCD}(q \leq q_\text{obs}))$, which is required to be less than 0.05 for
a 95\% confidence level (\CL) of exclusion. Because of the large number of events in the low-\mjj
range, which constrain the systematic uncertainties, we obtain 2--30\% better observed limits on the
BSM scales and masses compared to the limits obtained using the method in the predecessor of this
search reported in Ref.~\cite{Sirunyan:2017ygf}, where the nuisance parameters were marginalized
rather than profiled.
\par}

In the limit calculations, not all \mjj ranges are included in the likelihoods;
only those that improve the expected limits by more than 1\% are used.
We use mass bins with $\mjj>3.6\TeV$ for the CI models, $\mjj>4.2\TeV$
for the ADD models, and $\mjj>4.8\TeV$ for the QBH models.  For the DM mediators, we use mass bins
that cover the \mjj range of 0.5\MEDIMASS--1.2\MEDIMASS.  The exclusion limits on the BSM models are
determined using detector-level \CHI distributions and theoretical predictions at detector level.
By using the detector-level \CHI distributions, each bin of the \CHI distributions can be modeled by
a Poisson likelihood function, while at particle level, the unfolded data distributions have
correlations among the dijet mass bins.  As a cross-check, the limits are also determined for the
case where the unfolded \CHI distributions, approximated by Poisson likelihood functions, and
particle-level theoretical predictions are used in the limit extraction procedure.  The resulting observed
limits on the BSM scales and masses are found to be more stringent than those determined at detector
level by 1--10\%, depending on the model.  The agreement of the data with QCD predictions is
quantified by calculating $P_\text{QCD}(q < q_\text{obs})$ for each mass bin separately.  The
largest excess is found in the first data point of the $>$6.0\TeV mass bin,
with a significance of 1.8 standard deviations.
When combining mass bins in the various QCD+BSM models under study, the largest significances are
found to be 2.7--2.8 standard deviations for the QCD+DM model with $\MEDIMASS= 4.5$--6\TeV and
$\QCOUP=1.0$.

\begin{figure}[htb]
\centering
\includegraphics[width=\FigThreeWidth]{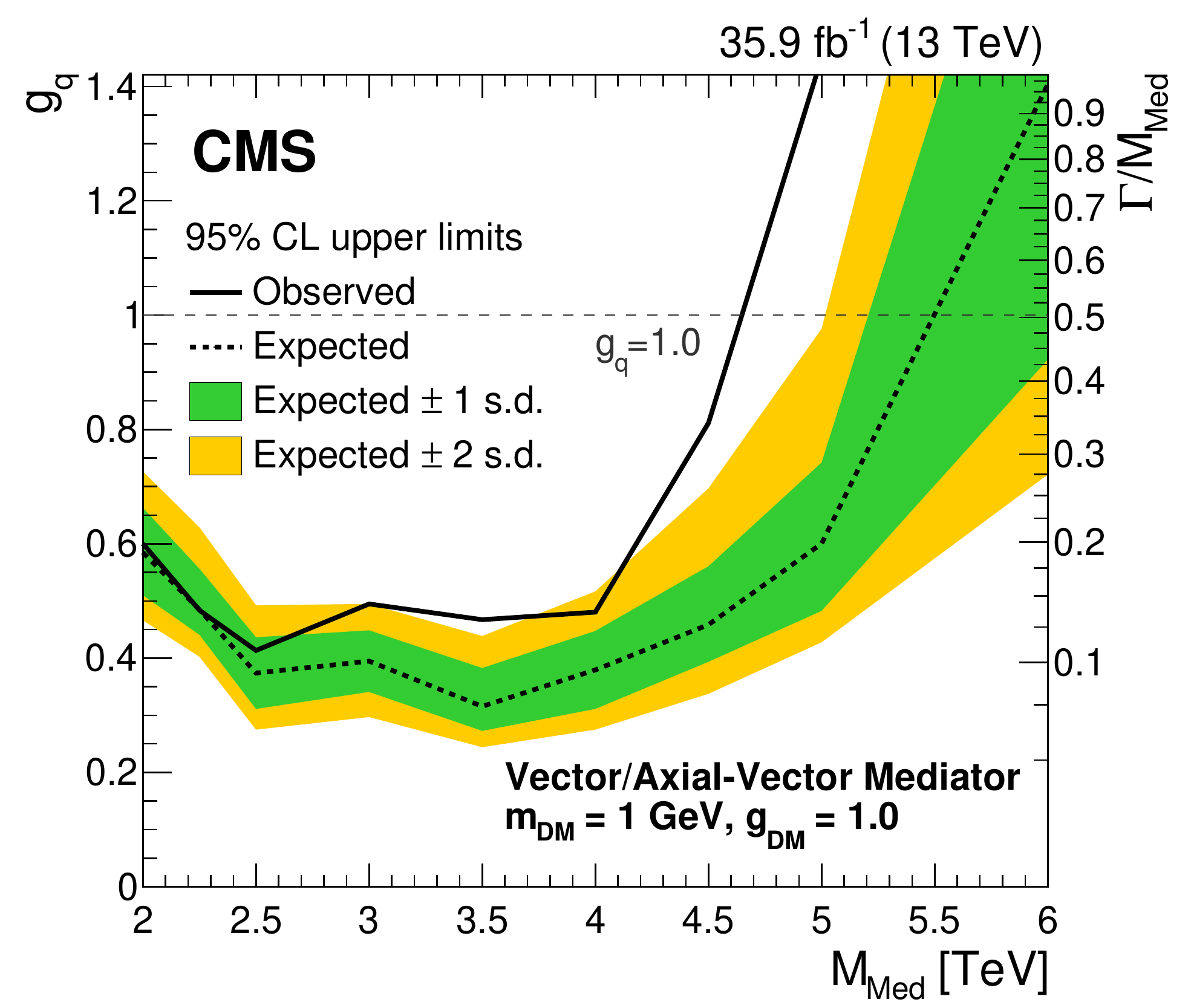}
\caption{
The 95\% \CL upper limits on the quark coupling \QCOUP, as a function of mass, for an axial-vector or vector DM mediator
with $\DMCOUP=1.0$ and $\DMMASS=1\GeV$.  The observed limits (solid), expected limits (dashed) and
the variation in the expected limit at the 1 and 2 standard deviation levels (shaded bands) are shown.  A dotted
horizontal line shows the coupling strength for a benchmark DM mediator with $\QCOUP=1.0$.  The
corresponding limits on the width of the mediators are shown on the vertical axis on the right-hand
side of the figure.}
\label{fig:DMcoupling}
\end{figure}

Figure~\ref{fig:DMcoupling} shows the 95\% \CL upper limits on \QCOUP as a function of the mass of
the vector or axial-vector DM mediator with $\DMCOUP=1.0$ and $\DMMASS=1\GeV$.  The corresponding
limits on the width of the mediators are shown on the vertical axis on the right-hand side of
Fig.~\ref{fig:DMcoupling}.  The degradation of the limits below $\MEDIMASS=2.5\TeV$ and above
$\MEDIMASS=4\TeV$ can be explained as follows.  For resonance masses below the lower \mjj boundary
of the analysis at 2.4\TeV, the acceptance increases rapidly as a function of resonance mass (\eg,
from 1.4\% at $\MEDIMASS=2\TeV$ to 16\% at $\MEDIMASS=2.5\TeV$, for $\QCOUP=0.5$), resulting in the
improvement of the limit on \QCOUP as a function of resonance mass.  For large values of resonance mass
and width (\eg, for $\MEDIMASS>4\TeV$ and $\QCOUP>0.5$), the mediator is primarily produced
off-shell with a mass less than the \mjj boundary of the analysis at 2.4\TeV.  The acceptance for
high resonance masses thus decreases as a function of resonance width (\eg, for $\MEDIMASS=5\TeV$,
from 25\% at $\QCOUP=0.5$ to 8\% at $\QCOUP=1.5$), resulting in the fast deterioration of the limit
on \QCOUP at high resonance masses.  The observed limit above 5\TeV is at $\RelWidth \geq 1$,
thus in a region where the simplified model of a mediator particle is no longer valid.
For \MEDIMASS between 2.0 and 4.6\TeV, this search excludes couplings $1.0 \leq \QCOUP \leq 1.4$,
which are not accessible via dijet resonance searches.

\begin{figure}[htb]
\centering
\includegraphics[width=\FigFourWidth]{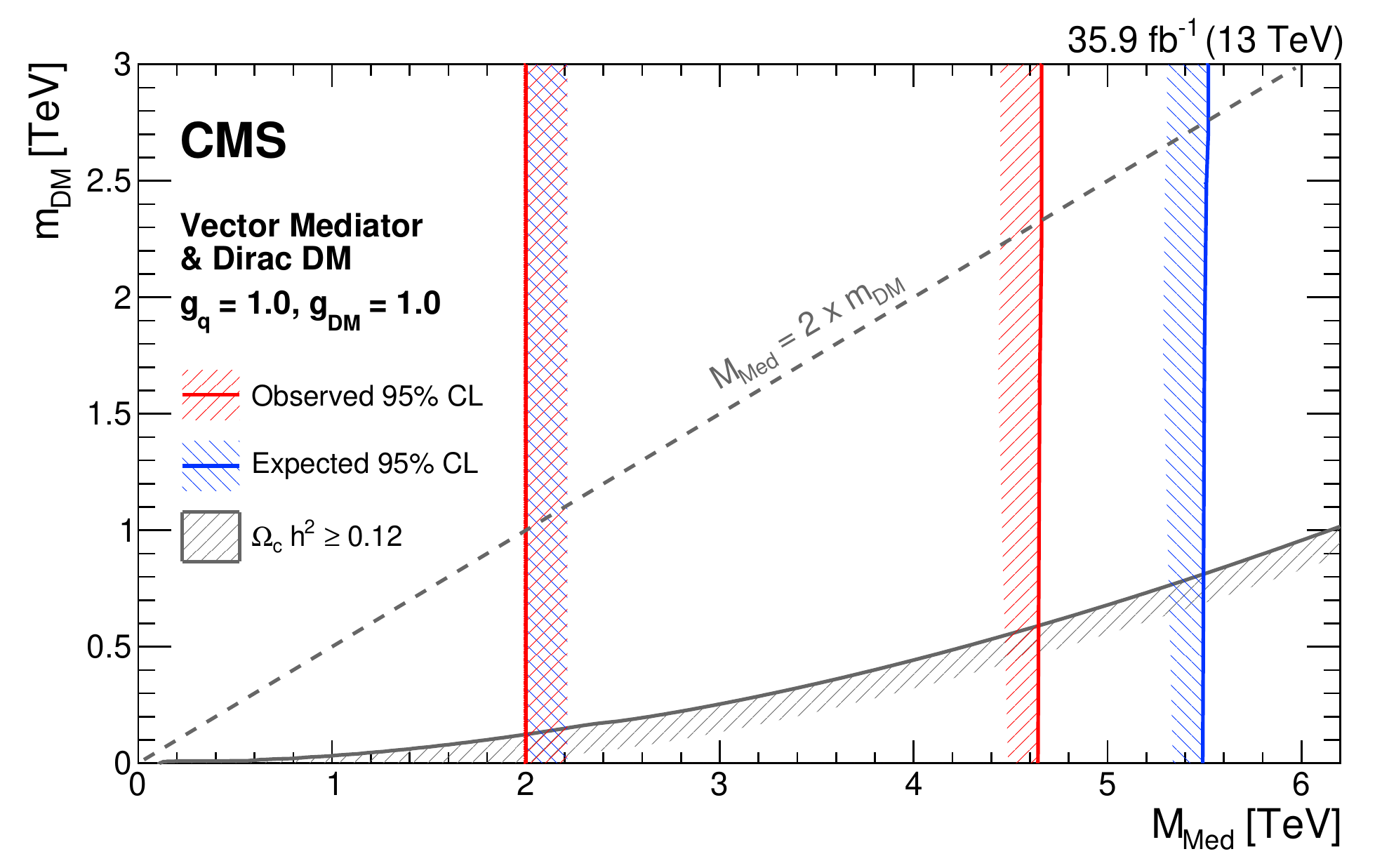}\\
\includegraphics[width=\FigFourWidth]{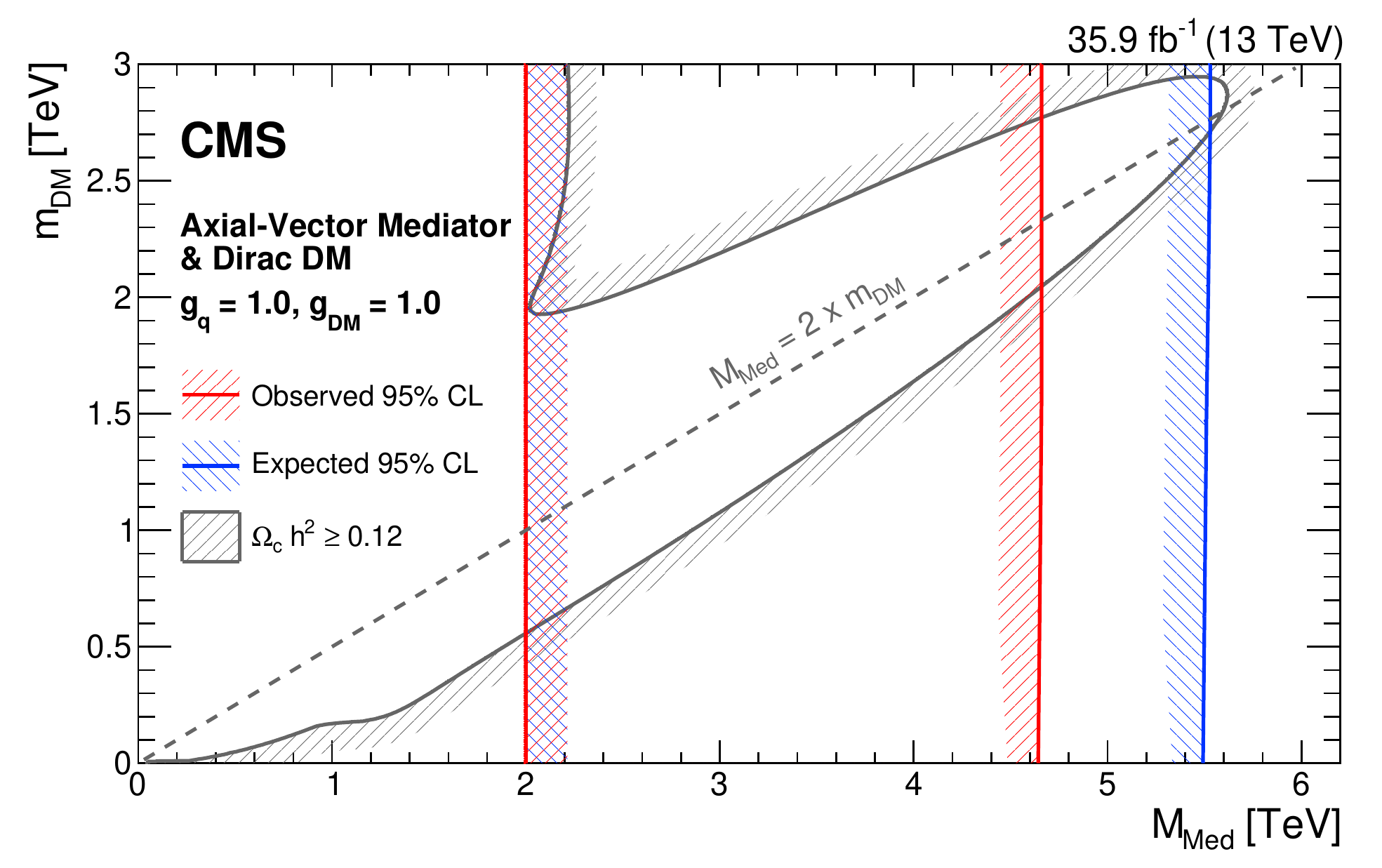}
\caption{
The 95\% \CL observed (red) and expected (blue) excluded regions in the plane of \DMMASS
and \MEDIMASS, for a vector mediator (upper) and an axial-vector mediator (lower) for a DM benchmark
model with $\DMCOUP=\QCOUP=1.0$. These are compared to constraints from the cosmological relic
density of DM (gray) determined from astrophysical measurements~\cite{Ade:2015xua}, using
\MADDM.  In the hatched area, DM is overabundant.  The observed and expected lower bounds
for \MEDIMASS overlap with each other.}
\label{fig:DMmass}
\end{figure}

The limits for \MEDIMASS at arbitrary \DMMASS and \DMCOUP can be calculated based on the fact that
at fixed mediator production cross sections, changes in the width of the DM decay channel
will lead to changes in the width of the quark decay channel.
For the models with $\QCOUP=1.0$, $\DMCOUP=1.0$, and $2\DMMASS<\MEDIMASS$, in which the total width
of the mediator is dominated by the width of the quark decay channel due to the large number of
possible quark flavors and colors, the exclusion range for \MEDIMASS has little
dependence on \DMMASS. For the models with $2\DMMASS\geq\MEDIMASS$, the width of the DM decay channel is
assumed to be zero. The resulting exclusion regions for vector and axial-vector mediators with
$\QCOUP=1.0$ and $\DMCOUP=1.0$ in the \DMMASS and \MEDIMASS plane are shown in
Fig.~\ref{fig:DMmass}.

The observed and expected exclusion limits at 95\% \CL on different CI, ED, QBH, and DM models obtained in this
analysis are listed in Table~\ref{tab:limits}.  The observed limits are less stringent
than the expected limits because of the upward fluctuation in the measured distributions compared to the
theoretical predictions.  The limits on all models are more stringent than those obtained from data
collected by CMS in 2015~\cite{Sirunyan:2017ygf}.

\begin{table*}[htb!!!]
\centering
\topcaption{
Observed and expected exclusion limits at 95\% \CL for various CI, ADD, QBH, and DM models.  The 68\%
ranges of expectation for the expected limit are given as well.  For the DM vector mediator,
couplings $\DMCOUP=1.0$, $\QCOUP\geq1$ and a DM mass of 1\GeV are assumed and a range of masses
instead of a lower limit is quoted.}
\label{tab:limits}
\begin{tabular}{lld{-2}D{,}{\pm}{-1}}
\multicolumn{2}{c}{Model}& \multicolumn{1}{c}{Observed lower limit (\TeV)} & \multicolumn{1}{c}{Expected lower limit (\TeV)} \\
\hline\\[-2ex]
CI & $\Lambda_{\mathrm{LL/RR}}^{+}$ & 12.8 & 14.6\, , \,0.8\\
&$\Lambda_{\mathrm{LL/RR}}^{-}$ & 17.5 & 23.5\, , \,3.0\\
&$\Lambda_{\mathrm{VV}}^{+}$ & 14.6 & 16.4\, , \,0.8\\
&$\Lambda_{\mathrm{VV}}^{-}$ & 22.4 & 30.7\, , \,3.7\\
&$\Lambda_{\mathrm{AA}}^{+}$ & 14.7 & 16.5\, , \,0.8\\
&$\Lambda_{\mathrm{AA}}^{-}$ & 22.3 & 30.6\, , \,3.8\\
&$\Lambda_{\mathrm{(V-A)}}^{+}$ & 9.2 & 11.5\, , \,1.0\\
&$\Lambda_{\mathrm{(V-A)}}^{-}$ & 9.3 & 11.8\, , \,1.1\\[\cmsTabSkip]
ADD & $\Lambda_\mathrm{T}$ (GRW) & 10.1 & 11.4\, , \,0.9 \\
& \MCUTOFF (HLZ) $\NEXTRA=2$ & 10.7 & 12.4\, , \,1.0 \\
& \MCUTOFF (HLZ) $\NEXTRA=3$ & 12.0 & 13.6\, , \,1.1 \\
& \MCUTOFF (HLZ) $\NEXTRA=4$ & 10.1 & 11.4\, , \,0.9 \\
& \MCUTOFF (HLZ) $\NEXTRA=5$ & 9.1 & 10.3\, , \,0.8 \\
& \MCUTOFF (HLZ) $\NEXTRA=6$ & 8.5 & 9.6\, , \,0.8 \\[\cmsTabSkip]
QBH & \QBHMASS (ADD $\NEXTRA=6$) & 8.2 & 8.5\, , \,0.4\\
& \QBHMASS (RS $\NEXTRA=1$)  & 5.9 & 6.3\, , \,0.7\\[\cmsTabSkip]
DM & Vector/Axial-vector \MEDIMASS  & \multicolumn{1}{c}{2.0\,--\,4.6} & \multicolumn{1}{c}{\hspace{10pt}2.0\,--\,5.5}\\
\end{tabular}
\end{table*}

\section{Summary}

A search has been presented for physics beyond the standard model, based on normalized dijet angular
distributions obtained in 2016 from proton-proton collisions at the
LHC.  The data sample corresponds to an integrated luminosity of \intlumi.
The angular distributions, measured over a wide range of dijet
invariant masses, are found to be in agreement with the predictions of
perturbative quantum chromodynamics. The results are used to set 95\% confidence level lower limits on the
contact interaction scale for a variety of quark compositeness models, the ultraviolet cutoff in
models of extra spatial dimensions, the minimum mass of quantum black holes, and the mass and
couplings in dark matter models.  For the first time, lower limits between 2.0 and 4.6\TeV are set
on the mass of a dark matter mediator for (axial-)vector mediators, for the universal quark
coupling $1.0\leq\QCOUP\leq1.4$.  This region is not accessible through dijet resonance searches.  The lower
limits for the contact interaction scale $\Lambda$ range from 9.2 to 22.4\TeV.  The lower limits on
the ultraviolet cutoff in the Arkani--Hamed--Dimopoulos--Dvali model are in the range of 8.5--12\TeV, and
are the most stringent limits available. Quantum black hole masses below 8.2\TeV are excluded in the
model with six large extra spatial dimensions, and below 5.9\TeV in the Randall--Sundrum model with
a single, warped extra dimension. To facilitate comparisons with the predictions of other models, the
angular distributions, corrected to particle level, are published in HEPData.

\clearpage

\begin{acknowledgments}
We congratulate our colleagues in the CERN accelerator departments for the excellent performance of the LHC and thank the technical and administrative staffs at CERN and at other CMS institutes for their contributions to the success of the CMS effort. In addition, we gratefully acknowledge the computing centres and personnel of the Worldwide LHC Computing Grid for delivering so effectively the computing infrastructure essential to our analyses. Finally, we acknowledge the enduring support for the construction and operation of the LHC and the CMS detector provided by the following funding agencies: BMBWF and FWF (Austria); FNRS and FWO (Belgium); CNPq, CAPES, FAPERJ, FAPERGS, and FAPESP (Brazil); MES (Bulgaria); CERN; CAS, MoST, and NSFC (China); COLCIENCIAS (Colombia); MSES and CSF (Croatia); RPF (Cyprus); SENESCYT (Ecuador); MoER, ERC IUT, and ERDF (Estonia); Academy of Finland, MEC, and HIP (Finland); CEA and CNRS/IN2P3 (France); BMBF, DFG, and HGF (Germany); GSRT (Greece); NKFIA (Hungary); DAE and DST (India); IPM (Iran); SFI (Ireland); INFN (Italy); MSIP and NRF (Republic of Korea); MES (Latvia); LAS (Lithuania); MOE and UM (Malaysia); BUAP, CINVESTAV, CONACYT, LNS, SEP, and UASLP-FAI (Mexico); MOS (Montenegro); MBIE (New Zealand); PAEC (Pakistan); MSHE and NSC (Poland); FCT (Portugal); JINR (Dubna); MON, RosAtom, RAS, RFBR, and NRC KI (Russia); MESTD (Serbia); SEIDI, CPAN, PCTI, and FEDER (Spain); MOSTR (Sri Lanka); Swiss Funding Agencies (Switzerland); MST (Taipei); ThEPCenter, IPST, STAR, and NSTDA (Thailand); TUBITAK and TAEK (Turkey); NASU and SFFR (Ukraine); STFC (United Kingdom); DOE and NSF (USA).

\hyphenation{Rachada-pisek} Individuals have received support from the Marie-Curie programme and the European Research Council and Horizon 2020 Grant, contract No. 675440 (European Union); the Leventis Foundation; the A. P. Sloan Foundation; the Alexander von Humboldt Foundation; the Belgian Federal Science Policy Office; the Fonds pour la Formation \`a la Recherche dans l'Industrie et dans l'Agriculture (FRIA-Belgium); the Agentschap voor Innovatie door Wetenschap en Technologie (IWT-Belgium); the F.R.S.-FNRS and FWO (Belgium) under the ``Excellence of Science - EOS" - be.h project n. 30820817; the Ministry of Education, Youth and Sports (MEYS) of the Czech Republic; the Lend\"ulet (``Momentum") Programme and the J\'anos Bolyai Research Scholarship of the Hungarian Academy of Sciences, the New National Excellence Program \'UNKP, the NKFIA research grants 123842, 123959, 124845, 124850 and 125105 (Hungary); the Council of Science and Industrial Research, India; the HOMING PLUS programme of the Foundation for Polish Science, cofinanced from European Union, Regional Development Fund, the Mobility Plus programme of the Ministry of Science and Higher Education, the National Science Center (Poland), contracts Harmonia 2014/14/M/ST2/00428, Opus 2014/13/B/ST2/02543, 2014/15/B/ST2/03998, and 2015/19/B/ST2/02861, Sonata-bis 2012/07/E/ST2/01406; the National Priorities Research Program by Qatar National Research Fund; the Programa Estatal de Fomento de la Investigaci{\'o}n Cient{\'i}fica y T{\'e}cnica de Excelencia Mar\'{\i}a de Maeztu, grant MDM-2015-0509 and the Programa Severo Ochoa del Principado de Asturias; the Thalis and Aristeia programmes cofinanced by EU-ESF and the Greek NSRF; the Rachadapisek Sompot Fund for Postdoctoral Fellowship, Chulalongkorn University and the Chulalongkorn Academic into Its 2nd Century Project Advancement Project (Thailand); the Welch Foundation, contract C-1845; and the Weston Havens Foundation (USA).
\end{acknowledgments}

\bibliography{auto_generated}

\cleardoublepage \appendix\section{The CMS Collaboration \label{app:collab}}\begin{sloppypar}\hyphenpenalty=5000\widowpenalty=500\clubpenalty=5000\vskip\cmsinstskip
\textbf{Yerevan~Physics~Institute, Yerevan, Armenia}\\*[0pt]
A.M.~Sirunyan, A.~Tumasyan
\vskip\cmsinstskip
\textbf{Institut~f\"{u}r~Hochenergiephysik, Wien, Austria}\\*[0pt]
W.~Adam, F.~Ambrogi, E.~Asilar, T.~Bergauer, J.~Brandstetter, E.~Brondolin, M.~Dragicevic, J.~Er\"{o}, A.~Escalante~Del~Valle, M.~Flechl, M.~Friedl, R.~Fr\"{u}hwirth\cmsAuthorMark{1}, V.M.~Ghete, J.~Grossmann, J.~Hrubec, M.~Jeitler\cmsAuthorMark{1}, A.~K\"{o}nig, N.~Krammer, I.~Kr\"{a}tschmer, D.~Liko, T.~Madlener, I.~Mikulec, E.~Pree, N.~Rad, H.~Rohringer, J.~Schieck\cmsAuthorMark{1}, R.~Sch\"{o}fbeck, M.~Spanring, D.~Spitzbart, A.~Taurok, W.~Waltenberger, J.~Wittmann, C.-E.~Wulz\cmsAuthorMark{1}, M.~Zarucki
% ...
\vskip\cmsinstskip
% ...
1:~Also at~Vienna~University~of~Technology, Vienna, Austria\\
% ...
\end{sloppypar}
\end{document}